\documentclass[draft=false,twocolumn,showpacs,preprintnumbers]{aa}
\usepackage{natbib,twoopt}
\usepackage{graphicx,units,multirow,subfigure,color,amsmath,amssymb,epsfig}
\usepackage[varg]{txfonts}
\usepackage[usenames,dvipsnames,svgnames,table]{xcolor}
\bibpunct{(}{)}{;}{a}{}{,} 

\definecolor{light-gray}{gray}{0.95}
\definecolor{dark-gray}{gray}{0.4}

\def\vectheta{\boldsymbol{\theta}}

\usepackage[pdftex,             
            breaklinks=true,%
            colorlinks=true,%
            pdfauthor={Ghelfi et al.},%
            pdftitle={Non-parametric determination of H and He IS fluxes from cosmic-ray data}%
           ]{hyperref}
\makeatletter
\newcommandtwoopt{\citeads}[3][][]{\href{http://adsabs.harvard.edu/abs/#3}{\def\hyper@linkstart##1##2{}\let\hyper@linkend\@empty\citealp[#1][#2]{#3}}}
\newcommandtwoopt{\citepads}[3][][]{\href{http://adsabs.harvard.edu/abs/#3}{\def\hyper@linkstart##1##2{}\let\hyper@linkend\@empty\citep[#1][#2]{#3}}}
\newcommandtwoopt{\citetads}[3][][]{\href{http://adsabs.harvard.edu/abs/#3}{\def\hyper@linkstart##1##2{}\let\hyper@linkend\@empty\citet[#1][#2]{#3}}}
\newcommandtwoopt{\citealpads}[3][][]{\href{http://adsabs.harvard.edu/abs/#3}{\def\hyper@linkstart##1##2{}\let\hyper@linkend\@empty\citealp[#1][#2]{#3}}}
\newcommandtwoopt{\citealtads}[3][][]{\href{http://adsabs.harvard.edu/abs/#3}{\def\hyper@linkstart##1##2{}\let\hyper@linkend\@empty\citealt[#1][#2]{#3}}}
\newcommandtwoopt{\citeyearads}[3][][]{\href{http://adsabs.harvard.edu/abs/#3}{\def\hyper@linkstart##1##2{}\let\hyper@linkend\@empty\citeyear[#1][#2]{#3}}}
\newcommandtwoopt{\citeadsstar}[3][][]{\href{http://adsabs.harvard.edu/abs/#3}{\def\hyper@linkstart##1##2{}\let\hyper@linkend\@empty\citealp*[#1][#2]{#3}}}
\newcommandtwoopt{\citepadsstar}[3][][]{\href{http://adsabs.harvard.edu/abs/#3}{\def\hyper@linkstart##1##2{}\let\hyper@linkend\@empty\citep*[#1][#2]{#3}}}
\newcommandtwoopt{\citetadsstar}[3][][]{\href{http://adsabs.harvard.edu/abs/#3}{\def\hyper@linkstart##1##2{}\let\hyper@linkend\@empty\citet*[#1][#2]{#3}}}
\newcommandtwoopt{\citeyearadsstar}[3][][]{\href{http://adsabs.harvard.edu/abs/#3}{\def\hyper@linkstart##1##2{}\let\hyper@linkend\@empty\citeyear*[#1][#2]{#3}}}
\newcommandtwoopt{\citeauthoradsstar}[3][][]{\href{http://adsabs.harvard.edu/abs/#3}{\def\hyper@linkstart##1##2{}\let\hyper@linkend\@empty\citeauthor*[#1][#2]{#3}}}
\newcommandtwoopt{\citepthesis}[3][][]{\href{http://tel.archives-ouvertes.fr/docs/#3}{\def\hyper@linkstart##1##2{}\let\hyper@linkend\@empty\citep[#1][#2]{#3}}}
\newcommandtwoopt{\citetthesis}[3][][]{\href{http://tel.archives-ouvertes.fr/docs/#3}{\def\hyper@linkstart##1##2{}\let\hyper@linkend\@empty\citet[#1][#2]{#3}}}
\makeatother

\begin{document}

\input epsf
\title{Non-parametric determination of H and He interstellar fluxes from cosmic-ray data}
\titlerunning{Non-parametric determination of H and He IS fluxes}

\author{A. Ghelfi\inst{1}\thanks{\url{alexandre.ghelfi@lpsc.in2p3.fr}}
   \and F. Barao\inst{2}
   \and L. Derome\inst{1}
   \and D. Maurin\inst{1}\thanks{\url{david.maurin@lpsc.in2p3.fr}}
}


\institute{
  LPSC, Universit\'e Grenoble-Alpes, CNRS/IN2P3,
      53 avenue des Martyrs, 38026 Grenoble, France
  \and
  LIP, P-1000 Lisboa, Portugal
}

\date{Received / Accepted}

\abstract
{Top-of-atmosphere (TOA) cosmic-ray (CR) fluxes from satellites and balloon-borne experiments are snapshots of the solar activity imprinted on the interstellar (IS) fluxes. Given a series of snapshots, the unknown IS flux shape and the level of modulation (for each snapshot) can be recovered.} 
{We wish (i) to provide the most accurate determination of the IS H and He fluxes from TOA data alone, (ii) to obtain the associated modulation levels (and uncertainties) while fully accounting for the correlations with the IS flux uncertainties, and (iii) to inspect whether the minimal force-field approximation is sufficient to explain all the data at hand.} 
{Using H and He TOA measurements, including the recent high-precision AMS, BESS-Polar, and PAMELA data, we performed a non-parametric fit of the IS fluxes $J^{\rm IS}_{\rm H,~He}$ and modulation level $\phi_i$ for each data-taking period. We relied on a Markov chain Monte Carlo (MCMC) engine to extract the probability density
function and correlations (hence the credible intervals) of the sought parameters.} 
{Although H and He are the most abundant and best measured CR species, several datasets had to be excluded from the analysis because of inconsistencies with other measurements. From the subset of data passing our consistency cut, we provide ready-to-use best-fit and credible intervals for the H and He IS fluxes from MeV/n to PeV/n energy (with a relative precision in the range $[2-10\%]$ at 1$\sigma$). Given the strong correlation between $J^{\rm IS}$ and $\phi_i$ parameters, the uncertainties on $J^{\rm IS}$ translate into  $\Delta\phi\approx \pm 30$~MV  (at 1$\sigma$) for all experiments. We also find that the presence of $^3$He in He data biases $\phi$ towards higher $\phi$ values by $\sim 30$~MV. The force-field approximation, despite its limitation, gives an excellent ($\chi^2/$dof$=1.02$) description of the recent high-precision TOA H and He fluxes.}
{The analysis must be extended to different charge species and more realistic modulation models. It would benefit from the AMS-02 unique capability of providing frequent high-precision snapshots of the TOA fluxes over a full solar cycle.}


\maketitle

\section{Introduction}

H and He interstellar (IS) fluxes are the most abundant species in the cosmic radiation. The low-energy part  of their spectrum (MeV/n to GeV/n) is responsible for ionising the interstellar
medium (ISM) \citepads{1987A&A...179..277W,1994MNRAS.267..447N,1998ApJ...506..329W,2012MNRAS.425L..86N} and molecular clouds \citepads[e.g.][]{2009A&A...501..619P}. They also interact with the ISM to produce 
light LiBeB isotopes \citepads{1970Natur.226..727R,1971A&A....15..337M,2000PhR...333..365V,2012A&A...542A..67P} and nuclear $\gamma$-rays  \citepads{1975A&A....40...91M,1977ApJ...211L..19L,1979ApJS...40..487R,2002ApJS..141..523K,2004NewAR..48...99T}. Uncertainties on the low-energy IS flux shape affect all these quantities \citepads{2009ApJ...694..257I}. The high-energy part of the IS flux (from GeV/n to PeV/n) is involved in the secondary production of $\gamma$-rays, neutrinos, antiprotons, and positrons \citepads{2007ARNPS..57..285S}. In particular, the hint at an energy break at hundreds of GeV energy  \citepads{2010ApJ...714L..89A,2011Sci...332...69A,2014PhRvL.112o1103A}, now confirmed by the AMS-02 collaboration \citep{2015PhRvL.114q1103A,PhysRevLett.115.211101}, must be accounted for as it affects the number of secondaries created \citepads{2011PhRvD..83b3014D,2011MNRAS.414..985L}. All these observables underline the necessity of a description as accurate as possible of the H and He IS fluxes over a wide energy range. 

A standard approach is to rely on top-of-atmosphere (TOA) data and simultaneously fit the IS flux parameters and solar modulation parameters \citepads{1975ApJ...202..265G,2004ApJ...612..262C,2006AdSpR..37.1727O,2007APh....28..154S}\footnote{Variations on this approach are to use IS fluxes obtained from cosmic-ray propagation codes \citepads[e.g.][]{2011A&A...526A.101P}, and/or the sparser radial-dependent CR data \citepads{2000JGR...10527447B,2003JGRA..108.8039L,2009JGRA..114.2103W} in the context of more realistic modulation models.}. At the crossroad of cosmic-ray and solar physics, these data give a unique perspective on the IS fluxes and the 22-year modulation cycle related to solar activity. The difficulty is that we do not know which solar modulation model (and parameters) to apply, but these models can be tested with high-statistics TOA data. We restrict ourselves in the present work to the simple force-field approximation \citepads{1967ApJ...149L.115G,1968ApJ...154.1011G,1987A&A...184..119P,1998APh.....9..261B}, which, despite some limitations \citepads[e.g.][]{2004JGRA..109.1101C}, is still widely used in the literature owing to its simplicity---it depends on a single free parameter $\phi(t)$.

Uncertainties on the IS fluxes translate into uncertainties on the solar modulation parameters. For instance, dark matter interpretations of the antiprotons and positrons fluxes are sensitive to $\phi$ uncertainties \citepads[e.g.]{2014PhRvD..90h1301L,2015JCAP...09..023G}. In general, there is no consensus on the required modulation level for a given data set and no consistency (of method and assumptions) between the various evaluations of $\phi$ provided by different experimental teams \citepads[see the discussion in][]{2014A&A...569A..32M}, which is problematic. Also directly related to the correlation between the IS flux and modulation level is the difficulty of establishing reliable and consistent modulation time-series from ground-based detector count rates \citepads{2011JGRA..116.2104U} and/or from the concentration of cosmogenic radionuclides in ice cores \citepads{2003JGRA..108.1355W,2010JGRA..11500I20H}. As underlined in several studies, the use of several IS flux parametrisations \citepads{1975ApJ...202..265G,2000JGR...10527447B,2003JGRA..108.8039L,2003JGRA..108.1355W,2006AdSpR..37.1727O,2007APh....28..154S,2009JGRA..114.2103W} leads to shifts of these time series up to $\Delta\phi\sim \pm 200$~MV \citepads[e.g.][]{2015AdSpR..55..363M}. This motivates
us even more to find a better characterisation of the IS flux, modulation level, and of the correlations between these parameters. The recent publication of high-precision data from PAMELA \citepads{2014PhR...544..323A}, BESS-Polar \citepads{2015arXiv150601267A}, and AMS-02 \citep{2015PhRvL.114q1103A,PhysRevLett.115.211101} is also a strong incentive to repeat and improve on the procedure of extracting these quantities from TOA CR data.

In Sect.~\ref{sec:methodology} we describe the use of  spline fit functions (in lieu of less flexible parametrisations previously used in the literature) to achieve a non-parametric determination of the H and He fluxes\footnote{Throughout the paper, we refer indifferently to $p$ or H to denote the proton flux; see discussion in Appendix~\ref{app:contamination}.}. A simple $\chi^2$ analysis is then used to select the subset of CR TOA data passing a consistency criterion. In Sect.~\ref{sec:MCMC} we replace the $\chi^2$ analysis by a Markov chain Monte Carlo (MCMC) exploration of the parameter space to obtain the credible intervals (CIs) and correlations between IS flux and modulation parameters. We also compare our results to the recent low-energy Voyager data, which are considered to be a direct probe of the local IS fluxes \citepads{2013Sci...341..150S,2013arXiv1308.6598W,2013arXiv1308.4426W,2013arXiv1308.1895W}, and to the indirect observation of IS fluxes in the direction of molecular complexes from Fermi-LAT $\gamma$-ray data \citepads{2012PhRvL.108e1105N,2012PhRvD..86d3004K,2014A&A...566A.142Y,2015arXiv150705006S}, before concluding in Sect.~\ref{sec:conclusion}. The short Appendix~\ref{app:systematics} investigates possible systematic effects on the $\phi$ determination that are due to deuteron and $^3$He contamination in H and He fluxes (\ref{app:contamination}), or when using TOA data obtained from long data-taking periods (\ref{app:time-average}).

\section{Methodology}
\label{sec:methodology}

To perform the analysis, the modulation model, the parametrisation of the IS flux, and the set of CR TOA data used for the minimisation must be specified. In this section, we present our setup, emphasising on the improvements made with respect to previous studies.

\subsection{Modulation: force-field approximation}
The simplest modulation model to link unmodulated (IS) to modulated (TOA) quantities is the force-field approximation  \citepads{1967ApJ...149L.115G,1968ApJ...154.1011G}:
\begin{eqnarray}
\label{eq:forcefield}
  \frac{E^{\rm TOA}}{A}&=&\frac{E^{\rm IS}}{A} - \frac{|Z|}{A} \phi\;, \\
  J^{\rm TOA} \left( E^{\rm TOA} \right)&=&
  \left( \frac{p^{\rm TOA}}{p^{\rm IS}} \right)^{2} \times J^{\rm IS}  \left( E^{\rm IS} \right), \nonumber
\end{eqnarray}
where $E$ is the total energy, $p$ the momentum, and $J\equiv dJ/dE_{k/n}$ is the differential flux per kinetic energy per nucleon $E_{k/n}$. This model has a single free parameter $\phi(t)$, whose dimension is rigidity.

\subsection{Analysis and $\chi^2$}
\label{sec:chi2}
The approach to break the degeneracy between $J^{\rm IS}$ and $\phi$ is to simultaneously fit $s$ different snapshots $\{t_1\dots,\, t_s\}$ of the same CR species $N_j$ and/or $n$ different CR species $\{N_1\dots,\,N_n\}$ at the same $t_i$. In the former approach, we benefit from sampling the same IS flux at the cost of one additional modulation parameter per snapshot: depending on the data precision and the periods used, ideally both high and low solar activity periods, the degeneracy is mostly lifted, although some significant uncertainties can remain \citepads[e.g.][]{2015AdSpR..55..363M}. In the second approach, the modulation level $\phi(t_i)$ is now the quantity sampled several times by different species, each new species requiring several additional shape parameters: the benefit is that species with different $Z/A$ (e.g. proton and helium) are differently modulated, even in the simple force-field approximation (see Eq.~\ref{eq:forcefield}). More generally, species with different charges (e.g. electrons and positrons) can probe different modulation models as a dependence on $Z$ is expected for example in drift modulation models \citepads[see][for a review]{2013LRSP...10....3P}.

The $\chi^2$ for a simultaneous fit over the TOA flux snapshots $t_i$, with several possibly measured species $N_j(i)$ at this $t_i$, and over all $E_k(i,j)$ energy bins measured, is given by
\begin{equation}
\label{eq:chi2}
\chi^2 = \sum_{t_i}\sum_{N_j(i)}\sum_{E_k(i,j)}\frac{\left(J^{\rm TOA}(J_j^{\rm IS},\phi_i, E_k)-{\rm data}_{ijk}\right)^2}{\sigma_{ijk}}\,,
\end{equation}
where the free parameters are the IS flux parameters for each species and the modulation parameters per TOA flux snapshot. The three cases considered in this study all involve fits over different snapshots $\{t_1\dots,\, t_s\}$ for 
\begin{itemize}
   \item $p$ data only,
   \item He data only, and
   \item $p$ and He data simultaneously.
\end{itemize}
The last option enables checking the consistency of the modulation levels derived with the values obtained from the separate species analysis (see Sect.~\ref{sec:MCMC}).

\subsection{Non-parametric IS flux: splines}
\label{sec:spline}

Fits of the IS fluxes in the literature are mostly based on simple power laws in total energy \citepads[e.g.][]{2006AdSpR..37.1727O,2014A&A...566A.142Y} or rigidity \citepads[e.g.][]{2007APh....28..154S,2014A&A...569A..32M}. Recently, to account for the high-energy break, broken power-laws were proposed \citepads{2011PhRvD..83b3014D,2011MNRAS.414..985L,2015PhRvL.114q1103A}. Whereas these parametrisations are flexible enough to describe the smooth behaviour above tens of GeV/n energies, the low-energy behaviour governed by a $\beta^a$ term (in front of the power-law) does not offer enough freedom to fit the log-parabola-shaped TOA flux. We find that pure and broken power laws in rigidity, kinetic energy, and total energy (i) give poor fits to the data for all parametrisations ($\chi^2/$dof$\sim 1.6-1.8$); (ii) do not allow assessing the relative merit of the different parametrisations, which complicates the task of providing a statistically meaningful description of the IS fluxes and their uncertainties.

To solve these problems, we used a spline function, which gives an excellent fit of the TOA data ($\chi^2/$dof$\lesssim 1$, see below) and also encompasses all the above parametrisations. A spline is a piecewise function defined by polynomials connecting at knots. The order of the spline is related to the highest order $n$ of the polynomial used, and smoothness is guaranteed by the fact that continuity of the spline and its $n-1$ derivatives are imposed. Here, we used a cubic spline ($n=3$) in $x=\log(R)$ and $y=\log(J^{\rm IS})$, with $R=pc/Ze$  the rigidity in GV. The $y$-values (flux) are the free parameters of the fit, while the number of knots $N$ and their $x$-position (rigidity) have to be set carefully to describe the shape of the data correctly. No structures are expected at low energy in the IS fluxes, therefore a small number of knots is enough to reproduce the smooth shape. Several knots are needed at and around the high-energy break position \citepads{2015PhRvL.114q1103A} to be able to match the structure. We find that at least six knots from 1 GV to 800 GV are necessary to provide a good description of the IS fluxes. The positions of the knots are the same for both H and He and are taken to be 
\begin{equation}
   \{R_0, \dots ,\,R_5\} = \{1,\,7,\,50,\,100,\,400,\,800\}~{\rm GV}.
\end{equation}
As a sanity check, we added at random $x$-positions additional knots and repeated the fitting analysis (described below). No difference in the best-fit spectrum being observed, we conclude that the six-knot third-order spline function provides a non-parametric determination of the IS fluxes.

\subsection{Data selection}
\label{sec:dataselection}
We retrieved TOA CR data for H and He from the cosmic-ray data base CRDB\footnote{\url{http://lpsc.in2p3.fr/crdb}} \citepads{2014A&A...569A..32M}. In principle, all the data should be used in the analysis. However, inconsistencies between different measurements are common, especially for the oldest datasets. To identify and then reject these inconsistent datasets, we performed a $\chi^2$ minimisation on all $p$ data,
see Eq.~(\ref{eq:forcefield}), for which we relied on the {\sc Minuit} minimisation package \citep{Minuit} from the {\sc Root} CERN libraries\footnote{\url{https://root.cern.ch}} \citepads{1997NIMPA.389...81B}. For each dataset, we then calculated an a posteriori distance between the data and the globally determined best-fit TOA flux for this set: 
\begin{equation}
\label{chi2_exp}
\chi^2_{\rm exp} = \left\{ \sum_{E_k}\frac{\left(J^{\rm TOA}(J_p^{\rm IS, best},\phi_{\rm exp}^{\rm best})-{\rm data}_{k}\right)^2}{\sigma_{k}} \right\} / n_{\rm data}\,.
\end{equation}
We then excluded all data with $\chi^2_{\rm exp}<2$. Strictly speaking, this quantity is just a distance and has no statistical bearing, but it nevertheless gives a good estimate of the goodness of fit of each dataset (see the values obtained in Table \ref{tab:Chi2_DataSelection}). This procedure accounts for modulation effects, hence allows checking the TOA data consistency over their full energy range. 

\begin{table}[!t]
\centering
\caption{List of proton and helium data tested and rejected ({\em italics}) for the analysis. The left column provides the name and date of the experiments; the second column gives (i) $\chi^2_{\exp}$ value (see Eq.~\ref{chi2_exp}) for proton fits using all the available data listed in this table and (ii) the same quantity, but only data for which the previous fit gives $\chi^2_{\rm exp}({\rm p})\leq2$; the third column is for  $\chi^2_{\rm exp}({\rm He})$ values, the cut sample now demanding that both $\chi^2_{\rm exp}({\rm p})\leq2$ and $\chi^2_{\rm exp}({\rm He})\leq2$.}
\label{tab:Chi2_DataSelection}
\begin{tabular}{llll}
\hline\hline
Experiment (date) & $\chi^2_{\rm exp}({\rm p})$&~& $\chi^2_{\rm exp}({\rm He})$\\
 & ~~~all $\rightarrow$ cut& & ~~~all $\rightarrow$ cut\\
\hline
AMS-01 (1998)         & 0.38  $\rightarrow$ 0.37 & & 7.6                      \\
AMS-02 (2011-2013)    & 1.4~~~$\rightarrow$ 1.2  & & 0.71  $\rightarrow$ 0.66 \\
{\em BESS93 (1993)}   & {\em 2.9}                & & {\em 2.5}                \\
BESS97 (1997)         & 0.12  $\rightarrow$ 0.11 & & 0.44  $\rightarrow$ 0.44 \\
BESS98 (1998)         & 0.45  $\rightarrow$ 0.43 & & 0.64  $\rightarrow$ 0.65 \\
BESS99 (1999)         & 0.24  $\rightarrow$ 0.23 & & 0.44  $\rightarrow$ 0.44 \\
BESS00 (2000)         & 1.1~~~$\rightarrow$ 1.0  & & 0.83  $\rightarrow$ 0.82 \\
{\em BESS-TEV (2002)} & {\em 4.5}                & & {\em 0.73}               \\
BESS-POLARI (2004)    & 1.5~~~$\rightarrow$ 1.6  & & 1.1~~~$\rightarrow$ 1.1  \\
BESS-POLARII (2007)   & 1.6~~~$\rightarrow$ 1.5  & & 0.46  $\rightarrow$ 0.49 \\
{\em CAPRICE98 (1998)}& {\em 6.9}                & & {\em \dots}              \\
{\em IMAX92 (1992)}   & {\em 2.6}                & & {\em 2.0}                \\
PAMELA (2006-2008)    & 0.27 $\rightarrow$ 0.26  & & {\em 4.5}                \\
PAMELA (2006/11)      & 0.34 $\rightarrow$ 0.35  & &\dots                     \\
PAMELA (2007/11)      & 0.28 $\rightarrow$ 0.29  & &\dots                     \\
PAMELA (2008/11)      & 0.22 $\rightarrow$ 0.24  & &\dots                     \\
PAMELA (2009/12)      & 0.09 $\rightarrow$ 0.09  & &\dots                     \\
\hline
\end{tabular}
\tablefoot{References for the data are AMS \citep{2000PhLB..490...27A,2015PhRvL.114q1103A,PhysRevLett.115.211101}, BESS \citepads{2002ApJ...564..244W,2007APh....28..154S,2015arXiv150601267A}, CAPRICE \citepads{2003APh....19..583B}, IMAX \citepads{2000ApJ...533..281M}, PAMELA \citepads{2011Sci...332...69A,2013ApJ...765...91A}.\vspace{-0.2cm}}
\end{table}

\begin{figure*}
\begin{center}
\includegraphics[width=\textwidth]{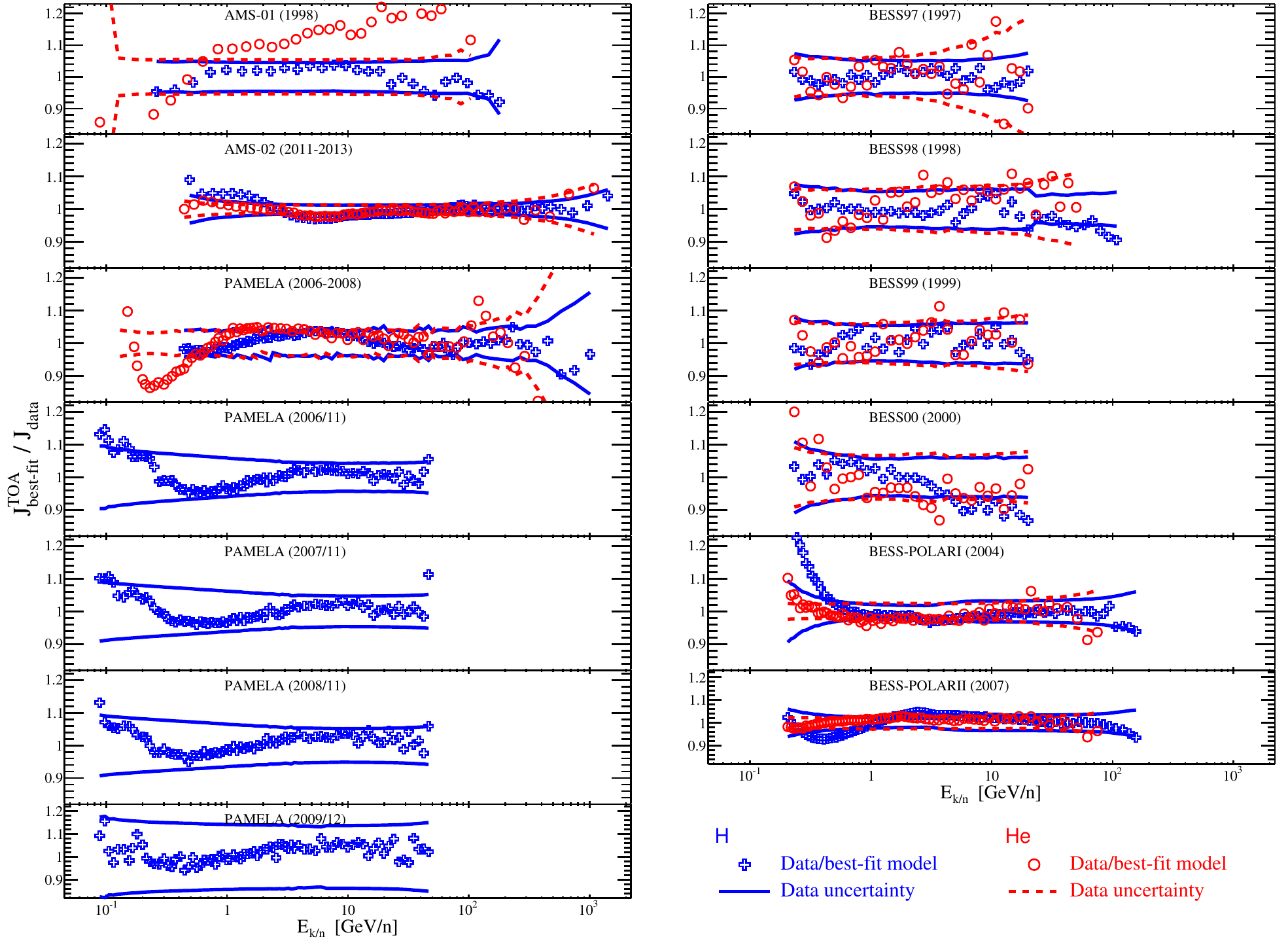}
\caption{Ratio of the best-fit model for $p$ (filled black circles) and He (empty red circles) to data for the experiments passing our selection (see Table~\ref{tab:Chi2_DataSelection}). The solid blue (dashed red) lines correspond to the uncertainties (statistical and systematics combined) on the $p$ (He) measurements. We note that the AMS-01 (top left panel) and PAMELA (2006-2008) He data (red empty circles; left panel, third row) are excluded based on their $\chi^2$ value (see Table \ref{tab:Chi2_DataSelection}) and are shown for illustration only.}
\label{fig:jtoa_over_jdata}
\end{center}
\end{figure*}

In practice, the procedure must be repeated several times because $\chi^2_{\rm exp}$ values are modified each time non-compatible datasets are removed, as shown in Table \ref{tab:Chi2_DataSelection}. Surprisingly, the AMS-01 and PAMELA (2006-2008) He data do not pass the cut. This is illustrated in Fig.~\ref{fig:jtoa_over_jdata}, which represents (symbols) the ratio between the best-fit model ($J^{\rm IS}_{\rm best}$ modulated by the associated $\phi_{\rm best}$) to TOA data. For good data, this ratio is mostly contained in the data error bars (solid and dashed lines for $p$ and He, respectively). AMS-01 He data (empty red circles in top-left panel) are an illustration of rejected data. We note that inconsistencies in the data at low energy may be indicative of a failure of the simple force-field approximation we used and not of underestimated systematics in the data, for instance. More realistic modulation models will be investigated in a forthcoming study. However, Fig.~\ref{fig:jtoa_over_jdata} shows for the very case of AMS-01 He data that the inconsistency with other datasets is not related to solar modulation as differences are at high energy.

From comparing the results from one experiment to another, we see no clear trend for a systematic bias towards lower or higher values of the fluxes. The only case for which a pattern in the energy dependence is observed is for the various PAMELA data
because they come from the same experimental setup and same analyses. This is not seen for BESS experiments, the configuration of which slightly changes between different flights.

We underline that neither Table~\ref{tab:Chi2_DataSelection} nor Fig.~\ref{fig:jtoa_over_jdata} show all the data rejected by the analysis. Moreover, some data were not considered for the MCMC analysis of Sect.~\ref{sec:MCMC} for the following reasons:
\begin{itemize}
  \item at low energy ($\lesssim$~GeV/n): several low-energy datasets passed the cut (several sets for ISEE-MEH \citepads{1986ApJ...303..816K} and Voyager \citepads{1983ApJ...275..391W}), but as they have no effect on the result (as checked from the $\chi^2$ minimisation analysis), we discarded them to avoid increasing the number of parameters (and runtime) in the MCMC analysis. 
  \item at intermediate energy: several datasets have a limited energy range with large error bars, hence no effect on the fit. For the same reason as above, they were not considered in the MCMC analysis.
  \item at high energy ($\gtrsim$~10 TeV/n): we recall that these data are not sensitive to solar modulation, but they determine the high-energy IS flux shape. We discarded them as the data are mostly inconsistent with one another (especially for He, see Fig.~\ref{fig:jis_comparison}).
\end{itemize}

\section{Markov chain Monte Carlo analysis}
\label{sec:MCMC}
On these data we used a Markov chain Monte Carlo (MCMC) algorithm
as implemented in the GreAT package \citepads{2014PDU.....5...29P}
to determine the probability density functions (PDF), correlations, and CIs for $\phi$ and IS flux parameters.

\subsection{MCMC algorithm}
\label{sec:MCMC_Principle}
The MCMC algorithm is based on the Bayes theorem that links the multidimensional PDF of the parameters to the likelihood function of the model and the prior on each parameter:
\begin{eqnarray}
\label{eq:Bayes}
\mathrm{PDF}(\vectheta| \vec{data})\propto \mathcal{L}\left(\vec{data}|\vectheta\right)\times \mathrm{P}\left(\vectheta\right)\,,
\end{eqnarray}
with $\vectheta$ corresponding to all IS flux and solar modulation parameters, and $\vec{data}$ the selected subset of experiments (see Sect.~\ref{sec:dataselection}). In the Bayesian interpretation of statistics, the function $\mathrm{P}\left(\vectheta\right)$ describes our prior knowledge on each parameter, and we took here a flat prior for all parameters. For the likelihood, with $\chi^2$ defined in Eq.~(\ref{eq:chi2}), we used
\[
   \mathcal{L}\left(\vec{data}|\vectheta\right) = \exp\left(-\frac{\chi^2(\vectheta,\vec{data})}{2}\right).
\]

The algorithm, based on random numbers, generates a chain $\vectheta_{i=1\dots N}$ of $N$ elements based on the following iterative process
(see e.g. \citealtads{2009A&A...497..991P}):
\begin{enumerate}
\item Randomly pick $\vectheta_0$ in the parameter space as the starting point of the chain; the current point is defined as $\vectheta_{\rm current} = \vectheta_0$.
\item Propose a new point $\vectheta_{\rm proposed}$ from a proposition function, here a multidimensional Gaussian, centred on $\vectheta_{\rm current}$.
\item Compute the probability of $\vectheta_{\rm proposed}$
 \[ \mathcal{P}\left(\vectheta_{\rm proposed}\right) = \mathcal{L}\left(\vec{data}|\vectheta_{\rm proposed}\right)\times \mathrm{P}\left(\vectheta_{\rm proposed}\right)\,,
 \]
and apply the Metropolis-Hastings criterion \citep{1993Neal,2003it...book....M}: $\vectheta_{\rm proposed}$ is added to the chain and becomes the new current point $\vectheta_{\rm current}$ with a probability 
   \[
    p=\min\left\{\frac{\mathcal{P}\left(\vectheta_{\rm proposed}\right)}{\mathcal{P}\left(\vectheta_{\rm current}\right)},1\right\}\;.
   \]
If $\vectheta_{\rm proposed}$ does not pass this criterion, $\vectheta_{\rm current}$ is added instead.
\item Repeat 2 until the chain has $N$ elements. 
\end{enumerate}

The chain is Markovian, meaning that the sampling of $\vectheta_{i+1}$ only depends on the value $\vectheta_i$. This implies a correlation between the consecutive values of $\vectheta$ in the chain. This correlation is corrected for by calculating the correlation length to select independent samples of $\vectheta^i$. A burning length is computed to estimate the number of steps needed to forget about the starting point, and these points are removed; for more details, see \citetads{2009A&A...497..991P} and \citetads{2014PDU.....5...29P}.
This procedure ensures that the multidimensional PDF of the parameters is correctly sampled. The outputs of the MCMC algorithm are 
\begin{itemize}
\item a natural marginalisation of the multidimensional PDF $\mathrm{PDF}(\vectheta| \vec{data})$ to access the 1D and 2D PDFs of the parameters,
and \item a vector of the model parameters that can be straightforwardly used to calculate CIs on quantities derived from the parameters (e.g. the IS fluxes for $p$ and He).
\end{itemize}

\begin{figure}
\includegraphics[width=\columnwidth]{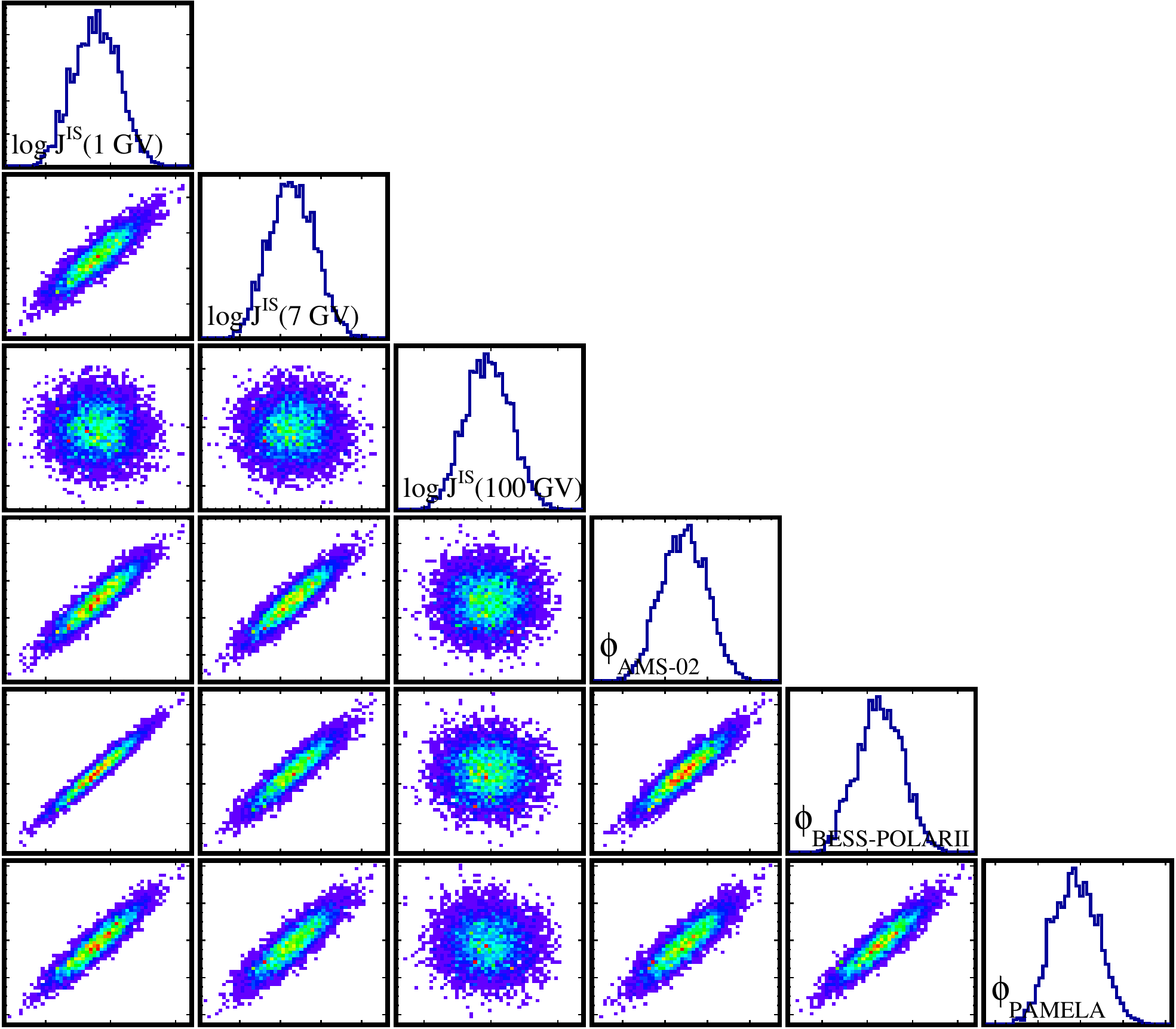}
\caption{PDF (diagonal) and 2D correlations (off-diagonal) plots for three selected knots $y=\log J^{\rm IS}(R)$ at position $\{R_0,\,R_1,\,R_3\}=\{1\,,7\,,100\}$~GV and three high-statistics datasets AMS-02, BESS-POLARII, and PAMELA (2006-2008): positive correlations are observed for all low-energy knots and datasets; knots above 100~GV show no correlation with any other parameter (the parameter distribution only depends then on the data uncertainties).}
\label{fig:pdf_jis_solmod}
\end{figure}

\subsection{Results of the MCMC analysis}
\label{sec:MCMC_Results}
\begin{figure}
\begin{center}
\includegraphics[width=\columnwidth]{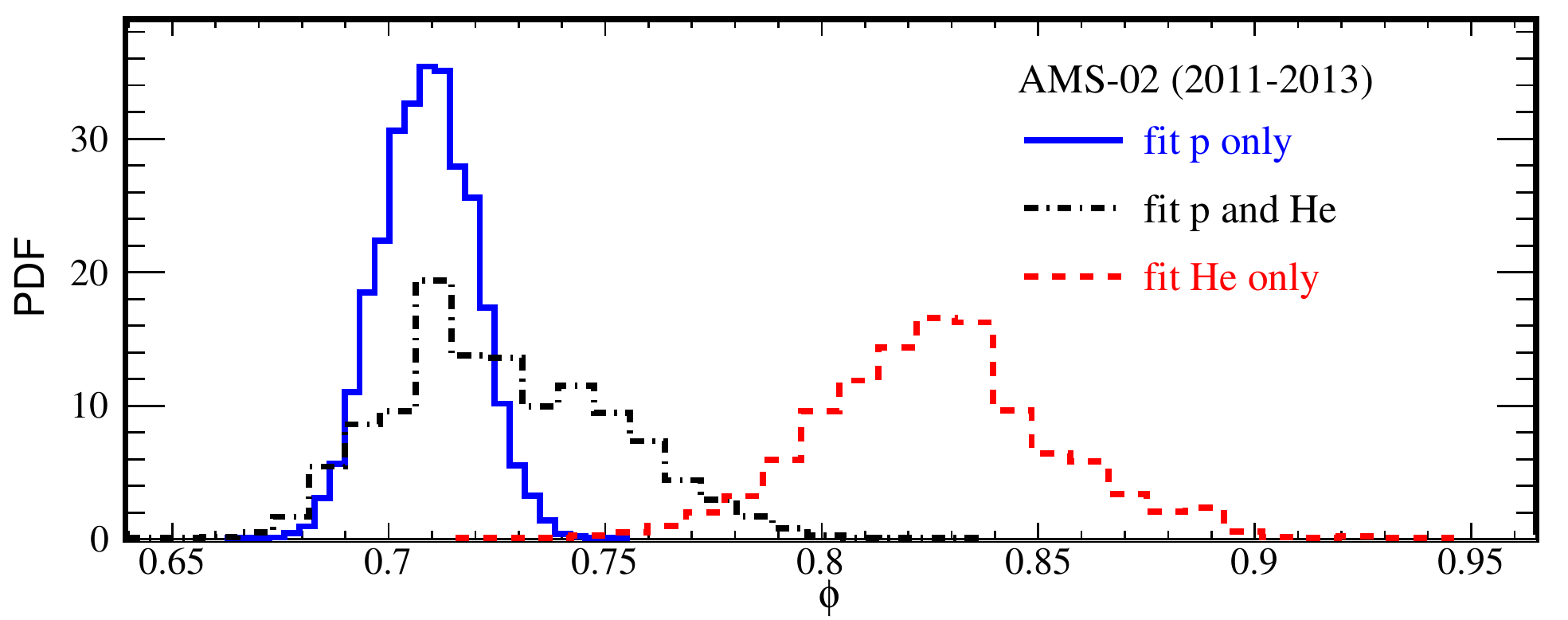}
\includegraphics[width=\columnwidth]{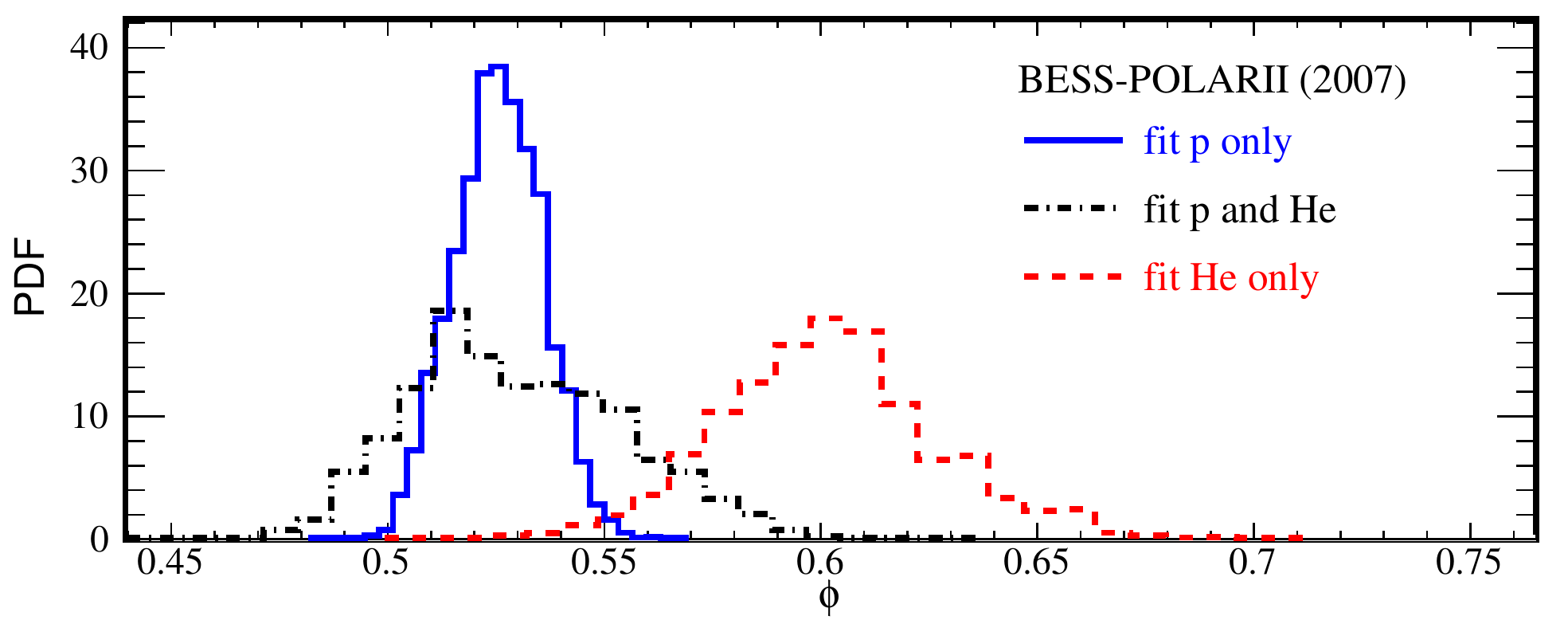}
\includegraphics[width=\columnwidth]{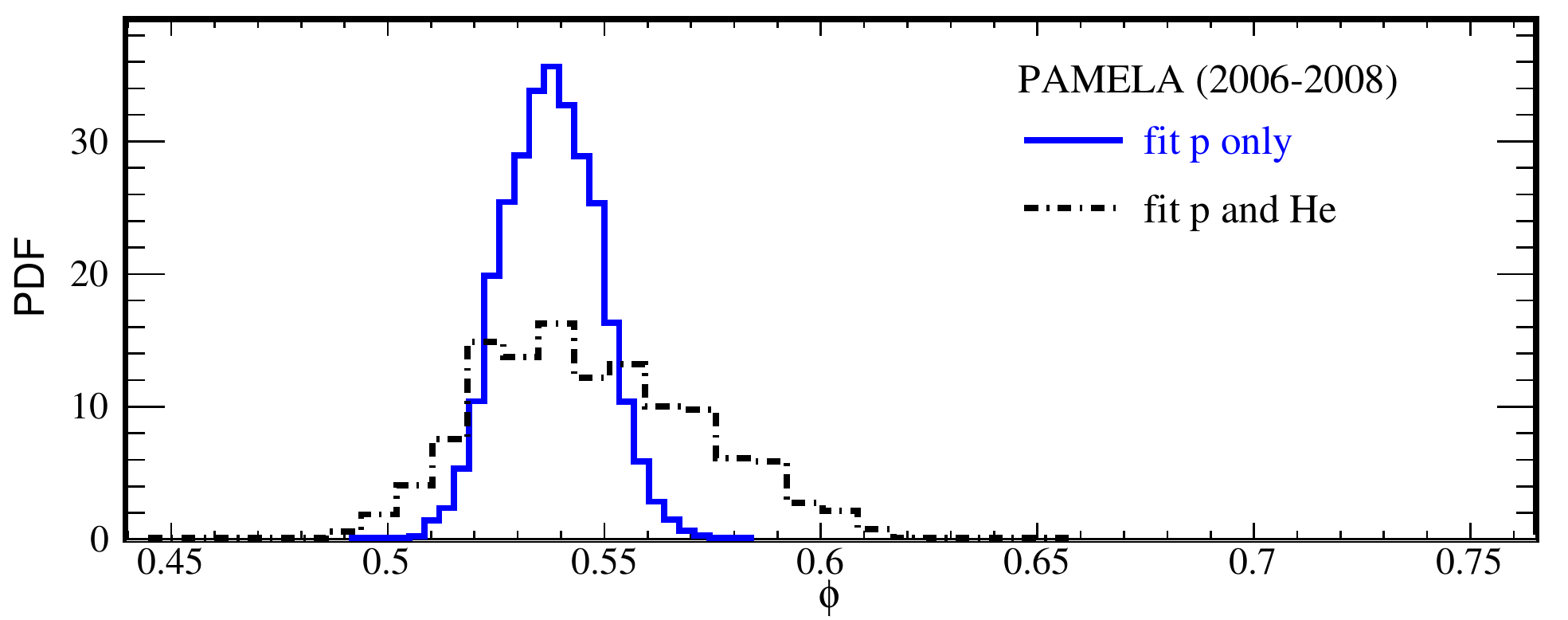}
\caption{PDF of the solar modulation level $\phi$ for the three most recent datasets with the highest statistics. We show the results for a fit on all selected data from Table~\ref{tab:Chi2_DataSelection}, for $p$ data alone (blue solid line), He data alone (red dashed line), and $p$ and He data simultaneously (black dash-dotted line).}
\label{fig:phi_values}
\end{center}
\end{figure}

\begin{figure}
\includegraphics[width=\columnwidth]{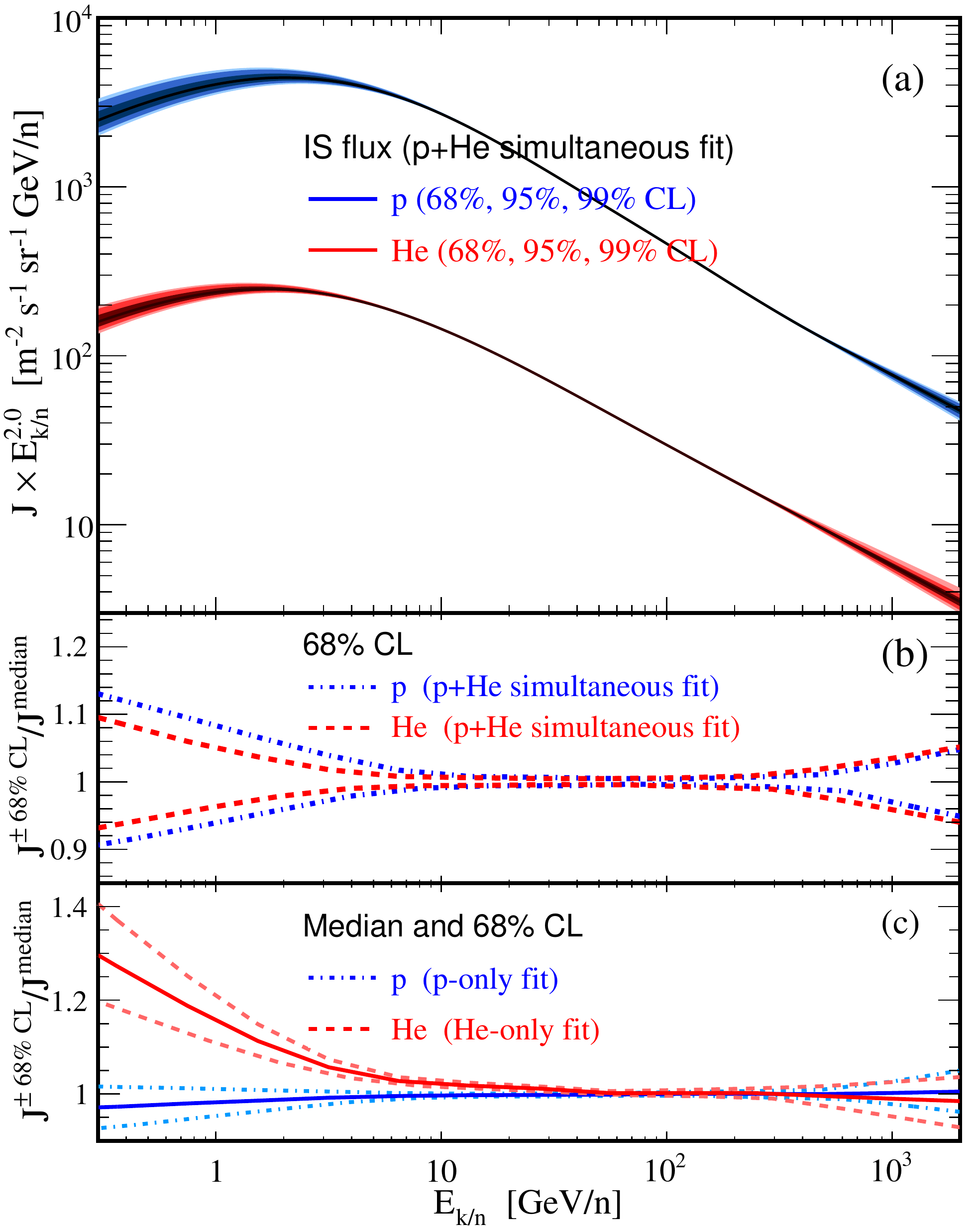}
\caption{Proton (blue) and helium (red) IS fluxes obtained from the various MCMC analyses. {\em Top panel (a):} 68\%, 95\%, and 99\% CIs (from darker to lighter shade) from the $p$+He analysis. The fluxes are multiplied by $E_{k/n}^2$ for illustration purposes. {\em Middle panel (b):} 68\% CI uncertainties for the $p$+He fluxes. {\em Bottom panel (c):} comparison of the $p$-only and He-only analyses to the $p$+He analysis. The solid lines correspond to the ratio of the median fluxes, the dash-dotted and dashed lines are for the 68\% CIs.}
\label{fig:jis}
\end{figure}

The marginalised PDFs for a selected subset of the IS flux and modulation parameters are presented in Fig.~\ref{fig:pdf_jis_solmod}. The PDFs for $\log J^{\rm IS}_{\rm p}$ and $\phi$ are close to Gaussian. The 2D PDFs show strong and expected correlations between the low-energy proton IS flux and the solar modulation levels: an increase of the IS flux must be balanced by an increase of the modulation level to recover the same TOA flux. The typical $\lesssim10\%$ uncertainty on (or dispersion between) the data at GeV/n energies (see Fig.~\ref{fig:jtoa_over_jdata}) translates into a similar uncertainty on $\phi$ values (see Fig.~\ref{fig:phi_values}) and on $\log J^{\rm IS}_{\rm p}$ at GV rigidities (see Fig.~\ref{fig:jis}). At high enough energy, as shown for the knot at $R=100$~GV (third parameter in Fig.~\ref{fig:pdf_jis_solmod}), the IS flux is no longer correlated to $\phi$. This reflects that solar modulation changes become irrelevant compared to data uncertainties (or dispersion between datasets).

As underlined in Sect.~\ref{sec:chi2}, the analysis can be performed separatley for proton and helium data or for both simultaneously (to probe the consistency of the derived $\phi$ values). The TOA data uncertainties propagate to the IS flux parameters, and then to $\phi$ values because of the correlations seen in Fig.~\ref{fig:pdf_jis_solmod}. Figure~\ref{fig:phi_values} shows the PDF of $\phi_{\rm AMS-02}$ (top), $\phi_{\rm BESS-POLARII}$ (middle), and  $\phi_{\rm PAMELA}$ (bottom)\footnote{We recall that PAMELA (2006-2008) He data did not pass the selection criterion (Table~\ref{tab:Chi2_DataSelection}), hence the corresponding result for He only is not reported in the bottom panel of Fig.~\ref{fig:phi_values}.}. The widths of the PDFs obtained in these three examples are representative of the width obtained for less accurate or lower-statistics experiments, indicating that the TOA data dispersion (between different experiments) dominates the TOA data uncertainties. We observe (at 1$\sigma$) typically $\Delta\phi_{\rm p}\approx \pm 10$~MV for the $p$-only analysis and $\Delta\phi_{\rm He}\approx\pm 30$~MV for both the He-only and $p$ and He simultaneous analyses. The larger uncertainty for He is the result of a larger scatter in He data than in $p$ data (He data uncertainties are similar to proton uncertainties, see in Fig.~\ref{fig:jtoa_over_jdata}), which is explained by the fact that He is less abundant and more difficult to measure. The central value for the $\phi_{\rm p+He}$ analysis is in between $\phi_{\rm p}$ and $\phi_{\rm He}$. We note that for the dataset where only $p$ data are available (e.g. PAMELA, bottom panel), the PDF obtained for the $p$ and He simultaneous analysis is also wider than for the $p$-only analysis because of the correlations between the parameters. Table~\ref{tab:phi} gathers the median and 68\% CI on the modulation level $\phi$ for the experiments selected in this analysis.

More important from Fig.~\ref{fig:phi_values} is the fact that the $\phi_{\rm p}$ and $\phi_{\rm He}$ values are not compatible at the 3$\sigma$ level. This could be an indication that the force-field approximation is rejected by the data. However, first, the best $\chi^2_{\rm best}/$dof values for the best-fit models found in the MCMC chains are 0.26 and 0.71 for these two analyses, possibly indicating an underestimation of the data uncertainties. For comparison, the $p$ and He simultaneous analysis gives a very good fit with $\chi^2_{\rm best}/$dof$=1.02$. Second, as shown in App.~\ref{app:contamination} on simulated data, the isotopic contamination of $^2$H and $^3$He leads to biased $\phi$ values, with an overshoot larger by typically $\sim 50$~MV for $\phi_{\rm He}$ than for $\phi_{\rm p}$. This allows us to reconcile the results of the $p$-only and He-only analyses. To reduce and eventually eliminate this bias, high-precision data (e.g. from AMS-02) are desired. 

\begin{table}[!t]
\centering
\caption{$\phi$ values obtained with the MCMC analysis for each experiment. The solar modulation is given as the median value computed from the PDF obtained fitting simultaneously $p$ and He (see Fig~\ref{fig:phi_values}. The errors correspond to a 68\% CI around the median value. The isotopic contamination from $^2$H and $^4$He in the data leads to an estimated bias (overshoot) of $\Delta\phi\approx 30$~MV (see Appendix~\ref{app:contamination}).}
\label{tab:phi}
\begin{tabular}{lr}
\hline\hline
Experiment (date)     & $\phi$ [MV]\\
\hline\vspace{0.05cm}
AMS-01 (1998)         &  $554^{+31}_{-26}$\\\vspace{0.05cm}
AMS-02 (2011-2013)    &  $724^{+31}_{-23}$\\\vspace{0.05cm}
BESS97 (1997)         &  $511^{+28}_{-22}$\\\vspace{0.05cm}
BESS98 (1998)         &  $606^{+31}_{-22}$\\\vspace{0.05cm}
BESS99 (1999)         &  $687^{+29}_{-23}$\\\vspace{0.05cm}
BESS00 (2000)         & $1309^{+32}_{-29}$\\\vspace{0.05cm}
BESS-POLARI (2004)    &  $776^{+30}_{-22}$\\\vspace{0.05cm}
BESS-POLARII (2007)   &  $527^{+29}_{-22}$\\\vspace{0.05cm}
PAMELA (2006-2008)    &  $544^{+30}_{-23}$\\\vspace{0.05cm}
PAMELA (2006/11)      &  $580^{+27}_{-21}$\\\vspace{0.05cm}
PAMELA (2007/11)      &  $505^{+25}_{-20}$\\\vspace{0.05cm}
PAMELA (2008/11)      &  $472^{+26}_{-19}$\\\vspace{0.05cm}
PAMELA (2009/12)      &  $409^{+23}_{-18}$\\
\hline
\end{tabular}
\end{table}

\subsection{Credible intervals (CI) on $J^{IS}$}

For each point $\vectheta$ in the MCMC chain, we can compute the associated $p$ and He IS fluxes at any energy. For each of these energies, we thus have a distribution of values from which we can calculate the PDF and CI; for rigidity values corresponding to the spline knots, the IS flux PDF at that point is directly the PDF from the MCMC analysis (see Fig.~\ref{fig:pdf_jis_solmod}). In Fig.~\ref{fig:jis} the top panel shows in various shades of blue (p) and red (He) the 68\%, 95\%, and 99\% CI on the IS flux (from the $p$+He simultaneous analysis). The middle panel  shows the 68\% contours divided by the median flux to emphasise the uncertainties: we have $\Delta J/J\lesssim 10\%$ at GeV/n, $\Delta J/J\lesssim 5\%$ above 1~TeV/n, and $\Delta J/J\lesssim 2\%$ in between. The uncertainties for $p$ and He are of the same order of magnitude. The bottom panel shows a comparison of the $p$-only and He-only analyses to the $p$+He simultaneous analysis (ratio of the median and 68\% contour to the median $p$+He flux). The median flux for $p$ is unchanged, and the $p$-only analysis leads to smaller uncertainties (compare the blue curves in panels (b) and (c)). The He fluxes are more sensitive to the analysis chosen, but they are compatible within their 95\% CIs (not shown). This last panel illustrates the fact that the fit is driven by $p$ data, so that the $p$+He combined analysis only affects He. As for the modulation parameters (see Sect.~\ref{sec:MCMC_Results}), it also illustrates the effect of correlations in the determination of the IS flux parameters because the $p$+He simultaneous analysis enlarges the uncertainties on the proton IS flux.

\begin{figure}
\includegraphics[width=\columnwidth]{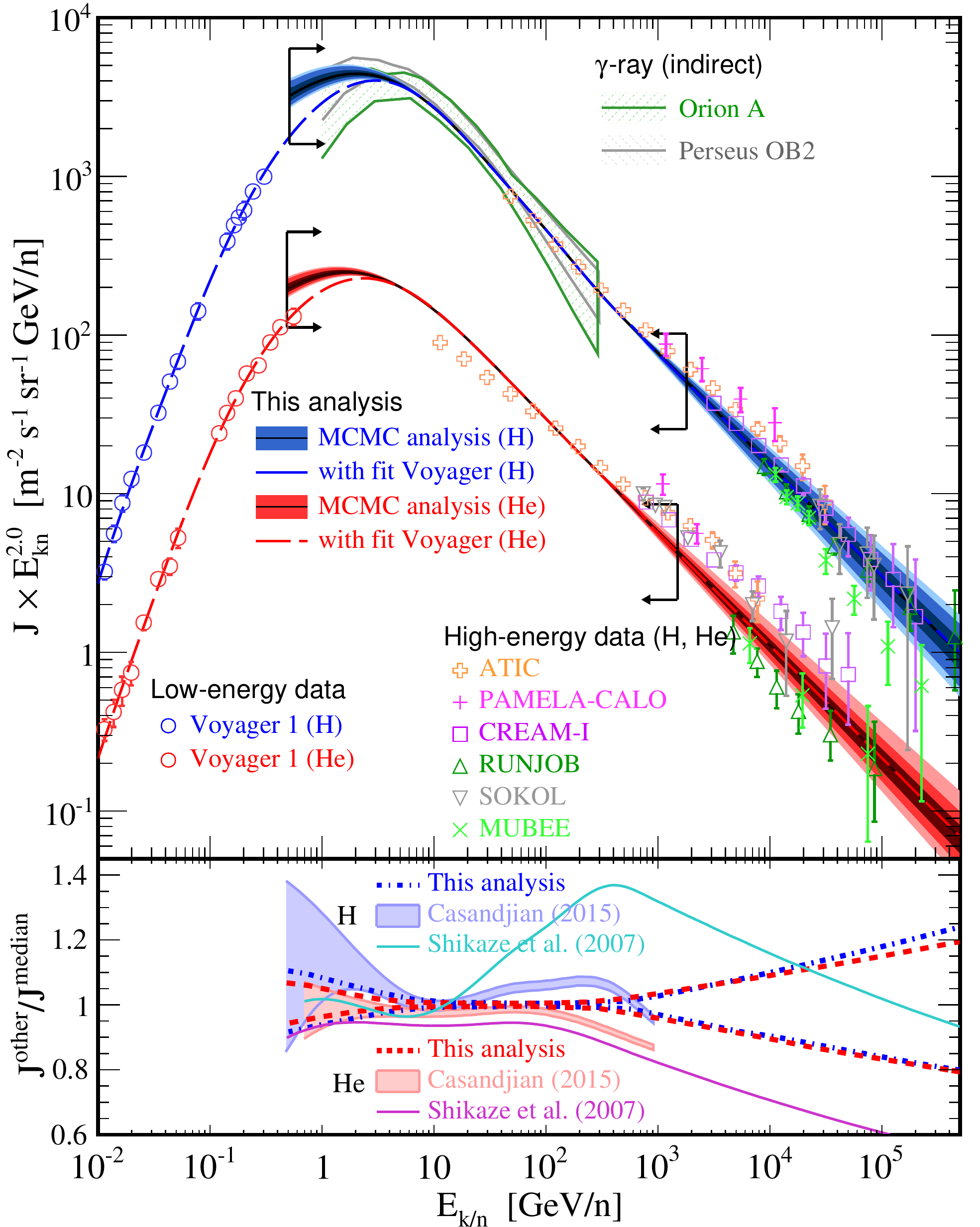}
\caption{Same as Fig.~\ref{fig:jis}, but compared to other direct or indirect IS flux determinations. {\em Top panel (a)}: green (Orion~A) and grey (Perseus~OB2) dashed areas are $\gamma$-ray derived limits from local giant molecular clouds using Fermi-LAT data \citepads{2014A&A...566A.142Y}; symbols are for low-energy Voyager~1 measurements \citepads{2013Sci...341..150S} and high-energy data. {\em Bottom panel (b)}: thick lines are from the 68\% CI of this analysis; shaded areas correspond to the 68\%CI from a likelihood analysis of Fermi-LAT $\gamma$-ray emissivity, and PAMELA (2006-2008) plus AMS-01 (1998) $p$+He+$e^+$+$e^-$ analysis \citepads{2015ApJ...806..240C}; chained lines are from a $\chi^2$-minimisation analysis of BESS data \citepads{2007APh....28..154S}.}
\label{fig:jis_comparison}
\end{figure}

\subsection{Comparison with other determinations}
\label{sec:comparisons}
The top panel of Fig.~\ref{fig:jis_comparison} shows comparisons with
\begin{itemize}
  \item a proton flux estimate from nearby giant molecular clouds in the Gould Belt. These fluxes are derived from $\gamma$-ray data, assuming that the interactions of CR with the ambient gas are fully responsible for the Fermi-LAT-observed fluxes \citepads{2014A&A...566A.142Y}. We find a very good match for the highest-emitting clouds (Orion A and Perseus OB2). Yang and collaborators compared their fluxes to the modulated PAMELA data (their Fig.~5 and~6) and found an excess below 10~GeV/n, which might be interpreted as CR that
are locally accelerated inside the cloud; when we compare this
to our IS flux instead, this turns into a deficit for most of the clouds that might be explained by increased energy losses and destruction of CRs in the clouds. 
  \item For the sake of completeness, we also compared our high-energy extrapolation (above the vertical arrow) to high-energy data that were not included in the fit: ATIC \citepads{2009BRASP..73..564P}, CREAM-I \citepads{2011ApJ...728..122Y}, JACEE \citepads{1998ApJ...502..278A}, MUBEE \citepads{1993ICRC....2...13Z}, PAMELA-CALO \citepads{2013AdSpR..51..219A}, RICH-II \citepads{2003APh....18..487D}, RUNJOB \citepads{2005ApJ...628L..41D}, and SOKOL \citepads{1993ICRC....2...17I}. Given the discrepancies between the datasets, our extrapolation is a fair estimate of the IS flux at these high energies. 
\end{itemize}
The bottom panel of Fig.~\ref{fig:jis_comparison} shows the ratio of various parametrisations of IS fluxes to our best-fit result, compared to the 68\% CIs on our IS fluxes (dashed and dash-dotted lines):
\begin{itemize}
  \item the best-fit and (shaded) contours of \citetads{2015ApJ...806..240C} were derived with a likelihood analysis based on PAMELA (2006-2008) and AMS-01 (1998) data (considering p, He, but also $e^+$ and $e^-$), plus the use of Fermi-LAT data on $\gamma$-ray emissivities. Overall, using more recent and more data yields smaller CIs. At high energies, the two results are not consistent within their 68\% CIs. At low energy, our smaller error bars may result from the use of a more flexible parametrisation (spline) than is the pure power-law used in \citetads{2015ApJ...806..240C}. 
  \item the best-fit fluxes (solid lines) of \citetads{2007APh....28..154S}
result from a fit to all BESS data in which the authors relied on a pure power-law. The high-energy feature is caused by the break in our parametrisation. In the energy range where both their fit and ours are not extrapolated, the largest difference between the two $p$ parametrisations is $\sim 30\%$ at 100 GeV/n, with their He flux systematically lower than ours.
\end{itemize}

\subsection{IS flux parametrisation with and without Voyager data}
It is also interesting to compare our flux derived from TOA-only data to the Voyager~1 data \citepads{2013Sci...341..150S}. The latter are believed to provide a direct measurement of the local IS (LIS) flux outside the solar cavity. However, the possibility of a small radial gradient in the outer heliosheath over hundreds of AU remains \citepads{2011ApJ...735..128S,2014ApJ...782...24K}. This must be balanced by indirect arguments \citepads{2014A&A...563A.108L} and simulation of this region \citepads{2014ApJ...793...18G,2015ApJ...808...82L,2015PhPl...22i1501Z} that support the hypothesis that Voyager has indeed reached the LIS.

To test the consequences of the two above alternatives, we added Voyager data (open circles in Fig.~\ref{fig:jis_comparison}) to the analysis. We then compared the results for the modulation values and IS fluxes obtained with and without Voyager data, either letting $\phi_{\rm Voyager}$ be a free parameter or enforcing it to be zero. In the former case, the best-fit with Voyager data is $\phi_{\rm Voyager}=65$~MV, without a modification of the IS fluxes. Conversely, if Voyager data are interstellar, we must enforce $\phi_{\rm Voyager}=0$: in that case, we observe (i) a decrease of $\sim 62$~MV on $\phi$ values for all other experiments, (ii) a worsened $\chi^2$/dof=1.27, and (iii) a lower best-fit IS flux at low energy (see long-dashed curves in Fig.~\ref{fig:jis_comparison}. We did not repeat the MCMC analysis because the (very small) Voyager uncertainties \citepads{2013Sci...341..150S} certainly do not include all the systematics.

Given the important role played by the H and He IS fluxes, we provide a simple implementation of our IS fluxes as a mixture of a log-polynomial and power-law extrapolation (spectral index $\tilde{c}_1$):
\begin{equation}
\log_{10}(J_{IS}) = 
  \begin{cases}
   \sum_{i=0}^{12} c_{i}\times \left(\frac{\log_{10}(E_{k/n})}{\log_{10}(800~{\rm GeV/n})}\right)^{i}
   \textrm{if $E_{k/n} < 800$~GeV/n};\\
   \tilde{c}_0 - \tilde{c}_1 \left(\frac{\log_{10}(E_{k/n})}{\log_{10}(800~{\rm GeV/n})}\right) \quad \textrm{otherwise}.
\label{eq:fit-poly}
   \end{cases}
\end{equation}
Table~\ref{tab:isfluxes} provides all theses coefficients for two cases (both obtained from the $p$+He simultaneous analysis):
\begin{itemize}
  \item the first five columns provide coefficients for the median and CIs from the MCMC analysis (without Voyager data), to be considered as a high estimate of the H and He IS fluxes (not valid below 400 MeV/n), and

  \item the last column provides the best-fit including Voyager data (with $\phi_{\rm Voyager}=0$), to be considered as a low estimate of the IS H and He fluxes.
\end{itemize}

\begin{table*}[!t]
\centering
\caption{Coefficients for H (top) and He (bottom) IS fluxes as parametrised by Eq.~(\ref{eq:fit-poly}). 
The first five columns correspond to the median and 1 and 2$\sigma$ CIs of the MCMC analysis (valid above 400 MeV/n). The last column corresponds to the best-fit fluxes when accounting for Voyager data (valid to lowest energy Voyager data).}
\label{tab:isfluxes}
\begin{tabular}{llrrrrrr}
\hline\hline
CR & Coeffs. & \multicolumn{5}{c}{MCMC analysis (without Voyager data)} & With Voyager \\
   &        & \multicolumn{1}{c}{$J_{\rm IS}^{\rm median}$} & \multicolumn{1}{c}{$J_{\rm IS}^{+1\sigma}$} & \multicolumn{1}{c}{$J_{\rm IS}^{+2\sigma}$} & \multicolumn{1}{c}{$J_{\rm IS}^{-1\sigma}$} & \multicolumn{1}{c}{$J_{\rm IS}^{-2\sigma}$} & \multicolumn{1}{c}{$J_{\rm IS}^{\rm best-fit}$}\\[2mm]

\multirow{13}{*}{H} & $c_{0}$ & 3.6088e+00 & 3.6427e+00 & 3.6753e+00 & 3.5824e+00 & 3.5626e+00 & 3.4617e+00  \\
 & $c_{1}$ & -5.1249e+00 & -5.2323e+00 & -5.3431e+00 & -5.0351e+00 & -4.9695e+00 & -4.1310e+00  \\
 & $c_{2}$ & -3.1517e+00 & -3.1232e+00 & -3.0871e+00 & -3.1816e+00 & -3.2058e+00 & -4.6403e+00  \\
 & $c_{3}$ & -1.9668e+00 & -1.8281e+00 & -1.6663e+00 & -2.1007e+00 & -2.1897e+00 & -1.4058e+00  \\
 & $c_{4}$ & 2.4595e+00 & 2.4752e+00 & 2.4718e+00 & 2.4542e+00 & 2.4545e+00 & -4.7537e+00  \\
 & $c_{5}$ & 1.8051e+00 & 1.6870e+00 & 1.5352e+00 & 1.9485e+00 & 2.0266e+00 & 8.5077e+00  \\
 & $c_{6}$ & -5.5338e-01 & -5.6684e-01 & -5.6738e-01 & -5.4072e-01 & -5.3201e-01 & 3.2637e+01  \\
 & $c_{7}$ & 2.3487e-01 & 2.0121e-01 & 1.6796e-01 & 2.3858e-01 & 2.6123e-01 & -2.8383e+01  \\
 & $c_{8}$ & -1.2658e+00 & -1.2431e+00 & -1.2024e+00 & -1.2894e+00 & -1.3050e+00 & -5.8203e+01  \\
 & $c_{9}$ & -2.1496e-01 & -1.9996e-01 & -1.8115e-01 & -2.4841e-01 & -2.5915e-01 & 4.8129e+01  \\
 & $c_{10}$ & -5.9272e-01 & -5.6908e-01 & -5.4286e-01 & -6.0271e-01 & -6.2241e-01 & 3.3946e+01  \\
 & $c_{11}$ & 2.9974e-01 & 3.0871e-01 & 3.0692e-01 & 2.9706e-01 & 2.8423e-01 & -2.9586e+01  \\
 & $c_{12}$ & 6.6289e-01 & 6.5647e-01 & 6.4893e-01 & 6.6907e-01 & 6.7734e-01 & 6.1683e-01  \\
 & $\tilde{c}_{0}$ & -3.7996e+00 & -3.7905e+00 & -3.7836e+00 & -3.8090e+00 & -3.8169e+00 & -3.7995e+00 \\
 & $\tilde{c}_{1}$ &  2.7040e+00 & 2.6737e+00 & 2.6415e+00 & 2.7354e+00 & 2.7718e+00 & 2.7040e+00  \\[3mm]
    
\multirow{13}{*}{He} & $c_{0}$ & 2.3742e+00 & 2.3953e+00 & 2.4147e+00 & 2.3583e+00 & 2.3461e+00 & 2.2784e+00  \\
 & $c_{1}$ & -5.3456e+00 & -5.4355e+00 & -5.5227e+00 & -5.2749e+00 & -5.2210e+00 & -4.5726e+00  \\
 & $c_{2}$ & -2.9866e+00 & -2.9022e+00 & -2.8139e+00 & -3.0600e+00 & -3.1233e+00 & -4.8650e+00  \\
 & $c_{3}$ & -1.3773e+00 & -1.2677e+00 & -1.1418e+00 & -1.4715e+00 & -1.5316e+00 & 3.9567e-01  \\
 & $c_{4}$ & 2.4858e+00 & 2.3740e+00 & 2.2262e+00 & 2.5976e+00 & 2.6921e+00 & -1.1578e+00  \\
 & $c_{5}$ & 1.3122e+00 & 1.2090e+00 & 1.0749e+00 & 1.4155e+00 & 1.4734e+00 & 4.9893e+00  \\
 & $c_{6}$ & -5.5501e-01 & -5.0560e-01 & -4.1371e-01 & -6.2188e-01 & -6.6670e-01 & 1.6511e+01  \\
 & $c_{7}$ & 1.2550e-01 & 1.1578e-01 & 1.0941e-01 & 1.2485e-01 & 1.2104e-01 & -2.0521e+01  \\
 & $c_{8}$ & -1.1544e+00 & -1.0715e+00 & -9.6435e-01 & -1.2320e+00 & -1.2943e+00 & -2.8367e+01  \\
 & $c_{9}$ & -1.7496e-01 & -1.6825e-01 & -1.6336e-01 & -1.8303e-01 & -1.8847e-01 & 3.1850e+01  \\
 & $c_{10}$ & -5.4768e-01 & -5.2429e-01 & -5.0422e-01 & -5.6130e-01 & -5.7343e-01 & 1.5000e+01  \\
 & $c_{11}$ & 2.5321e-01 & 2.0779e-01 & 1.4605e-01 & 2.9555e-01 & 3.3392e-01 & -1.7083e+01  \\ 
 & $c_{12}$ & 6.6468e-01 & 6.5963e-01 & 6.5218e-01 & 6.7276e-01 & 6.7868e-01 & 6.0486e-01  \\
 & $\tilde{c}_{0}$ & -4.9263e+00 & -4.9129e+00 & -4.9009e+00 & -4.9400e+00 & -4.9540e+00 & -4.9261e+00  \\
 & $\tilde{c}_{1}$ & 2.7141e+00 & 2.6910e+00 & 2.6653e+00 & 2.7448e+00 & 2.7693e+00 & 2.7140e+00  \\
\hline
\end{tabular}
\tablefoot{The parameters $c_{11}$ and $\tilde{c_0}$ are combinations of the other parameters. The continuity of the function and its first derivative at $800$~GeV/n enforces
$c_{12}=\frac{1}{12}\left(\sum_{i=1}^{11}i\times c_i + \tilde{c_1}\left(\log_{10}(800)\right)^2\right)$ and $\tilde{c_0}=\sum_{i=0}^{11} c_i$.}
\end{table*}

\section{Conclusion}
\label{sec:conclusion}

We have revisited the determination of IS fluxes and solar modulation parameters from TOA data alone. We took advantage of recent high-statistics experiments (AMS-02, BESS-Polar, PAMELA), we relied on a non-parametric fit of the IS fluxes (based on spline functions), and we used an MCMC to extract the PDF, CIs, and the correlation between the sought parameters. A preliminary step of the analysis was a consistency check that allowed rejecting some of the data. With the remaining data, we obtained reliable constraints on the $p$ and He IS fluxes in the region of GeV/n to several hundreds of TeV/n. These fluxes are very important for many applications in the literature (ISM ionisation, CR secondary production, etc.),
therefore we provide in Table~\ref{tab:isfluxes} ready-to use parametrisations based on our best-fit and IS flux contours with and without Voyager data. 

Correlations between IS flux parameters and solar modulation parameters were found to be important to estimate all the CIs properly. We studied $p$ and He separately or simultaneously to check the consistency of the modulation (most experiments measured $p$ and He data during the same period). Although the preferred $\phi$ values are slightly different in the separate analyses, the simultaneous $p$+He analysis gives a very good fit ($\chi^2/$dof$=1.02$) to all the data, with a 1$\sigma$ uncertainty of $30$~MV. As in many previous studies, H and He data are assimilated to pure $^1$H and $^4$He. However, we have shown that the presence of $^2$H and $^3$He leads to a $\sim 30$~MV positive bias in the $p$+He analysis, which is already similar to the systematic uncertainty. The bias is dominated by  the presence of $^3$He in He ($\sim50$~MV bias in the He-only analysis), therefore this contamination needs to be accounted for in future studies. 

From the data we conclude that the simple force-field approximation is effective in providing a good description (within the uncertainties) of the modulated $p$ and He fluxes at Earth. The situation may change because AMS-02 data have the capability and the statistics to provide monthly, weekly, or even daily average fluxes. This could be used to check the limitation of the force-field (or more evolved models, e.g. \citealtads{2015arXiv151108790C} and \citealtads{2016PhRvD..93d3016C}) over a full solar modulation cycle, in particular during a solar polarity change. Adding more species to the analysis, for example, $e^+$ and $e^-$ \citepads{2004ApJ...612..262C}, or even antiprotons, is the obvious next step to proceed with characterising IS CR fluxes and the solar modulation.

\paragraph{Note added.---} During the completion of this work, we became aware of a related study by Corti, Bindi, Consolandi and Whitman \citepads{2015arXiv151108790C}. It focuses on the IS proton flux and explores a modulation model beyond the force-field approximation. The data sets and methods both differ from those of our study, making the two analyses complementary. A comparison of their proton IS flux (obtained with the energy-dependent
solar modulation parameter) and ours (obtained in the force-field approximation) shows a very good agreement in the range 4~GV~--~1~TeV, but with different uncertainties.

\begin{acknowledgements}
We thank J.-M. Casandjian for providing the values of his IS flux determination, and C.~Combet for a careful reading of the manuscript. This work has been supported by the ``Investissements d'avenir, Labex ENIGMASS" and by the French ANR, Project DMAstro-LHC, ANR-12-BS05-0006. This study used the CC-IN2P3 computation center of Lyon.
\end{acknowledgements}

\appendix
\section{Systematics on $\phi$ determination}
\label{app:systematics}

\subsection{Effect of deuteron ( $^3$He) contamination of H (He) in $\phi$ determination}
\label{app:contamination}
Cosmic-ray experiments rarely achieve isotopic separation, even at low energy. Because solar modulation affects isotopes differently, see Eq.~(\ref{eq:forcefield}), the deuteron contamination of a few percent in H or the $\sim 20\%$ $^3$He contamination in He that peaks at GeV/n energy \citepads[e.g.][]{2012A&A...539A..88C} must be accounted for. 

To test the effect of isotopic contamination on the modulation levels, we proceeded in three steps:
\begin{enumerate}
  \item We estimated the interstellar fluxes for $^2$H and $^3$He running the same analysis as for $p$ and He data (see Sect.~\ref{sec:methodology}), that is:
  \begin{itemize}
    \item we retrieved $^2$H and $^3$He TOA data from CRDB~\citepads{2014A&A...569A..32M},
    \item for the IS flux determination, only three knots for deuterons at $\{0.8,\,3,\,7\}~{\rm GV}$ and two knots for $^3$He at $\{0.6,\,2\}~{\rm GV}$ are necessary, and
    \item the data passing the selection cut are ISEE~\citepads{1986ApJ...303..816K}, CAPRICE \citepads{1999ApJ...518..457B,2004ApJ...615..259P}, IMAX \citepads{2000AIPC..528..425D}, AMS-01 \citepads{2002PhR...366..331A,2011ApJ...736..105A}, and PAMELA \citepads{2013ApJ...770....2A}. The minimisation on these data provides the best-fit IS fluxes for $^2$H and $^3$He.
  \end{itemize}

  \item For each $p$ and He TOA data set, we simulated data based on the IS fluxes obtained from our analysis. The data points of each data set reflect the uncertainties of the experiment they comes from. Assuming that the above $^2$H and $^3$He IS fluxes are perfectly known, we modulated $J^{\mathrm{H}}-J^{^2\mathrm{H}}$  and $J^{\mathrm{He}}-J^{^3\mathrm{He}}$ with the corresponding solar modulation levels in Table~\ref{tab:phi}.
  
  \item We then repeated the fit and compared the obtained $\phi$ to those in Table~\ref{tab:phi}. We observe a systematic deficit for all modulation levels at the level of $\Delta\phi\in[24-33]$~MV for the $p$ +He simultaneous analysis, and $\Delta\phi\in[53-74]$~MV for the He-only analysis.
\end{enumerate}
We conclude that the presence of $^2$H and $^3$He (in H and He data, respectively) induces a non-negligible bias of $\sim 30$~MV in the determination of $\phi$, which is to be compared to the $\Delta\phi\sim \pm30$~MV systematic uncertainty from the MCMC analysis (see Fig.~\ref{fig:phi_values}). This bias is larger $\sim 65$~MV for the He-only  analysis, but smaller for the $p$-only analysis, providing a way to reconcile the discrepant values obtained in Fig.~\ref{fig:phi_values}.

\subsection{Effect of TOA flux long-time average on $\langle\phi\rangle$ determination}
\label{app:time-average}
PAMELA (2006-2008) and AMS-02 data are time averages over three years of the TOA fluxes. Assuming that the IS flux is perfectly known (taken to be the best-fit flux obtained in this paper), we compared two different calculations of the modulation value (all brackets below correspond to time averages):
\begin{itemize}
  \item $\phi_{\langle J^{\rm TOA} \rangle}$ calculated from the TOA flux $\langle J^{\rm TOA} \rangle$ obtained over a given data-taking period (e.g. the above-mentioned AMS-02 data), and
  \item $\langle\phi_{J^{\rm TOA}}\rangle$ calculated from the average of all modulation levels associated with $J^{\rm TOA}$ sampled over the whole data-taking period.
\end{itemize}

Solar modulation does not linearly transform IS fluxes, therefore
there is no reason for these two quantities to be equal. To estimate the effect of using one or the other approach, we proceeded as follows:
\begin{enumerate}
\item We used neutron monitor data to sample a realistic $\phi_i$ time series over the AMS-02 data-taking period,
\item we calculated for each time of the series the associated $J^{\rm TOA}_i$,
\item we calculated $\langle\phi\rangle=\sum_i \phi_i/N$ and $\langle J^{\rm TOA}\rangle =\sum_i J^{\rm TOA}_i/N$, and
\item we calculated $\phi_{\langle J^{\rm TOA} \rangle}$ fitting $\langle J^{\rm TOA}\rangle$ using AMS-02 errors.
\end{enumerate}

We find that $\langle J^{\rm TOA}(\phi_i)\rangle$ is within a few percent of $J^{\rm TOA}(\langle\phi_i\rangle)$, whereas  $\langle\phi\rangle$ and $\phi_{\langle J^{\rm TOA} \rangle}$ differ by 5 MV. To generalise this result, different three-year time periods were tested to probe the effect of stronger or different solar activity levels; the $\sim5$~MV difference remains. This is negligible compared to $\Delta\phi\sim \pm30$~MV obtained in Fig~\ref{fig:phi_values}.

\bibliographystyle{aa} 
\bibliography{nm_jis}
\end{document}